# ATOMISM VERSUS HOLISM IN SCIENCE AND PHILOSOPHY

*Vassilios Karakostas*[1]

[1]Department of Philosophy and History of Science
University of Athens, 157 71 Athens, Greece *

## ABSTRACT

The pros and cons of various forms of atomism and holism that are applicable both in physical science and today's philosophy of nature are evaluated. To this end, Lewis' thesis of Humean supervenience is presented as an important case study of an atomistic doctrine in philosophical thought. According to the thesis of Humean supervenience, the world is fragmented into local matters of particular fact and everything else supervenes upon them in conjunction with the spatiotemporal relations among them. It is explicitly shown that Lewis' ontological doctrine of Humean supervenience incorporates at its foundation the so-called separability principle of classical physics. In view of the systematic violation of the latter within quantum mechanics, it is argued that contemporary physical science posits non-supervenient relations over and above the spatiotemporal ones. It is demonstrated that the relation of quantum entanglement constitutes the prototypical example of a holistic, irreducible physical relation that does not supervene upon a spatiotemporal arrangement of Humean qualities, undermining, thereby, the thesis of Humean supervenience. Specifically, any relation of quantum entanglement among the parts of a compound system endows the overall system with properties which are neither reducible to nor supervenient upon any (intrinsic or extrinsic) properties that can possibly be attributed to each of its parts. It is concluded, in this respect, that the assumption of ontological reductionism, as expressed in Lewis' Humean doctrine, cannot be regarded as a reliable code of the nature of the physical world and its contents. It is proposed instead that — due to the undeniable existence of generic non-supervenient relations — a metaphysic of relations of a moderate kind ought to be acknowledged as an indispensable part of our understanding of the natural world at a fundamental level.

---

* Correspondence to: (E-mail: karakost@phs.uoa.gr)

## 1. Conceptions of Atomism and Holism in Physical Science

In the context of the atomism/holism issue in the natural sciences, if, broadly speaking, one adopts a strict bottom-up reasoning, starts thereby with the constituent parts of a system, maintains that these constituents have the properties that are characteristic of them independently of each other, and conceives a whole as a mere aggregate of its constituent parts, then one advocates atomism. In contrast, one recognizes the necessity of a holistic conceptualization if, roughly speaking, one claims that the constituents of a whole have, at least, some of the properties that are characteristic of them only within the whole. Although holism is frequently seen as the opposite of atomism, as the preceding general characterizations denote, nonetheless, atomism and holism may also be regarded as complementary viewpoints, in that, they both are needed to obtain a proper account of a given system.

Atomism in the natural sciences has historically functioned as the metaphysical background to classical physics. Within the latter all physical processes are completely described by a local assignment of basic physical magnitudes. The behavior, for instance, of a system of point particles under the action of forces is entirely determined by the assignment of particular values of position and momentum at each space-time point on their trajectories. As a result, a system in classical physics can always be analyzed into parts, whose states and properties determine those of the whole they compose (Section 3).

At the philosophical level of discourse, atomism, in accordance with Teller's (1989, p. 213) contemporary analysis, holds that the basic inhabitants of the natural world are distinct individuals, possessing qualitative, intrinsic properties and that all relations between individuals supervene on the possessed properties of the relata.[1] This means that if a, b, c, ... is a set of individual objects, then, any relational properties and physical relations holding among the relata a, b, c, ... supervene upon their qualitative, intrinsic properties. Intuitively, and for purposes of fixing the terminology to be used in the sequel, a property of an object-system is considered as being *intrinsic* (and hence *non-relational*) if the object has this property in and of itself, independently of the existence of other objects, and a property is regarded as being *qualitative* if its instantiation does not depend on the existence of any particular individual. For example, the property of owing a particular thing or being the father of a particular person are examples of non-qualitative, individual relational properties.[2]

Given, however, Teller's version of atomism, it is not clear whether spatial or spatiotemporal relations enjoy a privileged status. The question is the following: do spatiotemporal relations belong to the supervenience basis, namely, to the set of basic lower-level properties on which upper-level properties supervene or not? If one treats all relations alike, then, following Teller's dictates, spatiotemporal relations between physical objects seem to supervene upon their qualitative, intrinsic properties. But this, as will be argued in the sequel, is not quite valid even within classical physics. At any rate, one could provide a more general than Teller's rather restrictive definition of atomism by appropriately broadening the supervenience basis so as to include pertinent relational properties among the related parts.

Generally speaking, a supervenience basis is formed by the set of properties and relations of the parts upon which the properties and relations of the whole may or may not supervene. However hard is to precisely specify the supervenience basis, it is essential that one should avoid adding *global* properties or relations to this basis. In such a circumstance the thesis of atomism, as well as its antagonistic view of holism, would be rendered trivial. For, if one were permitted to consider *all* properties and relations among the parts, then these trivially would determine the properties of the whole they compose. Thus, only certain properties and relations can be allowed in the supervenience basis. The latter is expected to include, according to Healey (1991, p. 401), only "the qualitative, intrinsic properties and relations of the parts, i.e., the properties and relations that these bear in and of themselves, without regard to any other objects, and irrespective of any further consequences of their bearing these properties for the properties of any wholes they might compose". For instance, spatial or spatiotemporal relations seem to provide the appropriate kind of physical relations that are required in the supervenience basis in addition to the qualitative, intrinsic properties of the related parts, since the spatio-temporal relations can vary independently of the properties of their relata and hence are not supervenient upon them. At least within classical physics, spatial or spatiotemporal relations among physical systems are the only clear examples of physical relations required in the supervenience basis for specifying physical atomism: other intrinsic physical relations seem to supervene on them. In this respect, determination of the qualitative, intrinsic properties of the parts together with their spatial or spatiotemporal relations is sufficient to determine all the qualitative, intrinsic properties of the whole along with its causal structure of relations as governed by the forces between the individual parts (Section 3). If, then, one includes spatial, or more generally, spatiotemporal relations into the supervenience basis, we arrive at the following conception of physical atomism:[3]

> *Physical atomism as property determination*: A physical compound system is atomistic if and only if every qualitative, intrinsic property and relation of the whole system supervenes upon the qualitative, intrinsic properties of its constituent parts together with the spatiotemporal relations among the parts.

On the assumption that the qualitative, intrinsic properties of a system specify the system's overall state, then physical atomism as a thesis of property determination gives rise to the following notion:

> *Physical atomism as state determination*: A physical compound system is atomistic if and only if the whole state of the compound system supervenes upon the states of its constituent parts.

This conception is closely related to the separability principle of classical physics (Section 3). It is captured intuitively by the familiar fact that if one constructs a compound system or object by assembling its independently existing parts, then the physical properties of that system are completely determined by the properties of the parts and the way they have been combined so as to form the initial system of interest. This exemplification characterizes also, as a natural restriction to physics, David Lewis' recombination principle (Section 4.3) that has been forwarded alongside with his ontological doctrine of Humean Supervenience (Section 2).

While atomism was apparently legitimized by the undeniable empirical successes of classical physics, nonetheless, developments in the conceptual foundations of contemporary physics — especially quantum physics — have shown to resist atomism in favor of holistic considerations. Holism, as an emergent concept in the philosophy of quantum physics, arises from the behavior of *entangled* quantum systems and the associated conception of *non-* *separability*, the peculiar phenomenon of '*non-locality*' that appears in the so-called EPR-Bell-type correlations and the *non-individuality* of quantal entities (Section 4). All these prominent features of quantum theory cast severe doubts on the common view of the world as consisting of concrete, unchangeable, self-contained individuals, being localized in space-time, and existing independently of one another. These issues, in particular the notion of non-separability, all relate to holism (Sections 4.1 and 4.2).

As already mentioned in the introduction, holism, broadly speaking, is the thesis that certain properties of systems cannot be determined or explained by properties of their constituent parts. Instead, the system as a whole determines in an important way how the parts behave. Various epigrams have been used in various, diverse fields of knowledge in order to characterize holism in general terms, as for instance: 'the whole is more than the sum of its parts', 'the whole precedes its parts' or still 'the whole acts on its parts'. Ian Smuts, who coined the term holism in his 1926 book "Holism and Evolution", refers to the dialectics of the whole-parts relation as follows: "In the end it is practically impossible to say where the whole ends and the parts begin, so intimate is their interaction and so profound their mutual influence. In fact so intense is the union that the differentiation into parts and whole becomes in practice impossible, and the whole seems to be in each part, just as the parts are in the whole" (Smuts 1987/1926, p. 126).

Any evaluation, of course, of the significance of holism in physics must rest on an adequate clarification of this notion. To the extent that a holistic conceptualization is applicable to physics, it is only natural to be contextualized to physical properties of compound physical systems or objects. We are interested here at the extent to which the properties and relations pertaining to a whole system are determined by the properties and relations that may be admitted, in a non-trivial manner, to the supervenience basis of its basic physical parts. Thus, in juxtaposition to atomism, we arrive at the following thesis of physical holism:

> *Physical holism as a denial of property determination*: A physical compound system is holistic if and only if not all of the qualitative, intrinsic properties and relations of the whole system supervene upon the qualitative, intrinsic properties and relations in the supervenience basis of its constituent parts.

In view of the preceding specification, a property of a physical compound system is holistic if and only if the description of the property in question cannot be determined or deduced by a description of the properties of its constituent parts and their admissible relations in the supervenience basis. Holistic properties pertaining to a system can affect the behavior of its individual constituents in ways that cannot be specified exclusively by reference to properties those constituents possess independently of their membership in the whole.

All instances of property holism arise because a physical system may have such irreducible properties, namely, properties whose possession by that system is neither determined by nor supervenient upon the properties of its component subsystems. If a system has such an irreducible property at a particular time, it may be said to be in a holistic state at that time. Consequently, the thesis of physical holism as a failure of property determination is closely related to a form of holism in terms of a failure of state determination:

*Physical holism as a denial of state determination*: A physical compound system is holistic if and only if the whole state of the compound system is not supervenient upon the states of its constituent parts.

Given that the parts of a genuinely holistic system depend on each other in such a manner that some of their properties that are characteristic of them can only be acquired within the whole, the following specification of physical holism may be adopted as well:

*Physical holism as a dependence on the whole*: A physical compound system is holistic if and only if it is only the whole state of the compound system that leads to a complete determination of the state-dependent properties of its constituent parts and the relations among these parts (to the extent that these are determined at all).

Due to the parts-whole interdependence, properties of individual constituents within a holistic system can only be determined by relations they bear to one another with respect to the whole. In other words, the properties of the parts of a holistic system are primarily *relational*. The question of whether quantum mechanical systems indeed exhibit holistic features, in the light of the preceding broad specifications, is examined in Section 4. For the time being we only note that the aforementioned characterization of physical holism, adopting a top-down reasoning, is especially relevant in the framework of quantum theory where the very identity of microphysical parts depends upon their contexts and relationships within the whole. Now, we turn to a detailed investigation of the philosophical thesis of Humean supervenience, as a case study of an atomistic doctrine in philosophy, examining in particular its compatibility with the structure and findings of contemporary physics.

## 2. The Ontological Doctrine of Humean Supervenience

Over the last couple of decades David Lewis, the systematic late philosopher of the "Plurality of the Worlds", has been defending a metaphysical, ontological doctrine he calls 'Humean supervenience'. According to Lewis (1986a, p. ix), Humean supervenience is the doctrine — inspired by Hume, the great denier of necessary connections — that

> … all there is to the world is a vast mosaic of local matters of particular fact, just one little thing and then another. ... We have geometry: a system of external relations of spatio-temporal distances between points. Maybe points of space-time itself, maybe point-sized bits of matter or aether or fields, maybe both. And at those points we have local qualities: perfectly natural intrinsic properties which need nothing bigger than a point at which to be instantiated. For short: we have an arrangement of qualities. And that is all. There is no difference without difference in the arrangement of qualities. All else supervenes on that.

There is thus a distribution of local, qualitative, intrinsic properties whose instantiation requires no more than a spatiotemporal point. Between these points there exist spatio-temporal relations or occupation relations holding between point-objects and space-time points, or both. These relations do not supervene on the local, intrinsic properties of the related objects. The properties of everything else supervene on the distribution of intrinsic properties instantiated at space-time points or arbitrarily small regions of space-time. The world, therefore, is fragmented into local matters of particular fact and everything else supervenes upon them in conjunction with the spatiotemporal relations among them.

Since the notion of supervenience has been given a multiplicity of various technical definitions in order to help characterize, both in metaphysics and philosophy of mind, a wide variety of philosophical purposes (see, for instance, Savellos and Yalsin 1995; Kim 1993), a clarification of the leading notion of supervenience within Lewis' philosophical framework is required. Lewis defends the usefulness of the notion of supervenience, characterizing supervenience itself as a denial of independent variation. He notes:

> To say that so-and-so supervenes on such-and-such is to say that there can be no difference in respect of so-and-so without difference in respect of such-and-such. Beauty of statues supervenes on their shape, size and colour, for instance, if no two statues, in the same or different worlds, ever differ in beauty without also differing in shape or size or colour (1983, p. 358).

Lewis explicitly accommodates a dependence thesis with his view that "supervenience means that there *could* be no difference of the one sort without difference of the other sort", adding that "without the modality [indicated by 'could'] we have nothing of interest" (1986b, p. 15).

Thus, Lewis' conception of supervenience acquires in effect the form of a 'dependence – determination' relationship. The dependence aspect is that possessing a supervenient (or upper-level) property requires possessing some subvenient (or lower-level) property, whereas, the determination aspect is that possession of that subvenient property will suffice for possession of the supervenient property. Thus, on this conception of supervenience, possessing a supervenient property requires possessing some subvenient property whose possession suffices for the instantiation of the supervenient property in question. In other words, supervenient properties exist only because of the underlying or subvenient properties, and these are sufficient to determine (not necessarily explain) how the supervenient properties arise. It is in this respect that Lewis (1999, p. 29) claims that "a supervenience thesis is, in a broad sense, reductionist".

If, therefore, Lewis' core idea of supervenience is that once the underlying subvenient level is fixed, the upper or supervenient level is fixed as well, then, at least, it should be true that if two objects (e.g., the two statues in Lewis' quotation above) do not differ with respect to their subvenient properties, then they should also not differ with respect to the supervenient properties. Consequently, properties of, say, type-*A* are supervenient on properties of type-*B* if and only if two objects cannot differ with respect to their *A*-properties without also differing with respect to their *B*-properties. Thus, for example, global properties of a compound physical system, considered as a whole, supervene on local properties of its component parts if and only if there can be no relevant difference in the whole without a difference in the parts.

As a 'fairly uncontroversial' example of what supervenience can be like, Lewis (1986b, p. 14) offers the following:

> A dot-matrix picture has global properties — it is symmetrical, it is cluttered, and whatnot — and yet all there is to the picture is dots and non-dots at each point of the matrix. The global properties are nothing but patterns in the dots. They supervene: no two pictures could differ in their global properties without differing, somewhere, in whether there is or isn't a dot.

A dot-matrix pattern is of course supervenient upon the contingent arrangement of dots. The pattern is entailed by the intrinsic properties and distance relations among the dots. If the dots are there, the pattern is also there, in a manner that is entailed by the subvenient base of the dots. No two patterns could differ in their global properties (e.g., symmetry) without differing in their point-by-point arrangement of dots. Lewis, in formulating his thesis of Humean supervenience, takes the view that an analogous kind of entailment relation holds for the totality of all facts about the world, in the sense that all global matters of fact supervene upon a spatiotemporal arrangement of local base facts. In his words:

> Could two worlds differ … without differing, somehow, somewhere, in local qualitative character ? (1986b, p. 14)
> 
> The question turns on an underlying metaphysical issue. A broadly Humean doctrine (something I would very much like to believe if at all possible) holds that all the facts there are about the world are particular facts, or combinations thereof. This need not be taken as a doctrine of analyzability, since some combinations of particular facts cannot be captured in any finite way. It might better be taken as a doctrine of supervenience: if two worlds match perfectly in all matters of particular fact, then match perfectly in all other ways too (1986a, p. 111), [specifying that],
> 
> …the world has its laws of nature, its chances and causal relationships; and yet – perhaps! – all there is to the world is its point-by-point distribution of local qualitative character (1986b, p. 14).

In a world or part of a world where Lewis' doctrine of Humean supervenience holds, there are no necessities in and between the particular facts as such. All the particulars of such an aggregate whole are in their spatio-temporal existence both logically and nomologically independent of each other. Assuming the validity of Lewis' thesis of Humean supervenience, such a contingent spatiotemporal arrangement of local qualities provides an irreducible subvenient basis — what one may call a Humean basis — upon which all else supervenes. The Humean basis consists of intrinsic qualities instantiated locally at space-time points and external distance relations between them. There is no difference anywhere without a difference in the spatiotemporal distribution of intrinsic local qualities. Everything not itself in the Humean basis supervenes upon it. Hence, in Lewis' conception of Humean supervenience, spatio-temporal relations enjoy a privileged status; they are acknowledged as the *only fundamental external physical relations* that are required in the subvenient base. "In a world like ours", Lewis (1994, p. 474) remarks, "the fundamental relations are exactly the spatiotemporal relations: distance relations, both space-like and time-like … In a world like ours, the fundamental properties are local properties: perfectly natural intrinsic properties of points, or of point-sized occupants of points. All else supervenes on the spatiotemporal arrangement of local qualities".

## 3. Humean Supervenience and its Affinity to Classical Physics

Evidently, Lewis' formulation of Humean supervenience gives expression to some typical Humean claims like the 'looseness' and 'separateness' of things[4] in a manner that is cashed out in terms of intrinsic properties instantiated at distinct space-time points or arbitrarily small regions of space-time. It is important to realize in this respect that Lewis' Humean doctrine incorporates at its foundation the so-called separability principle of classical physics, a principle that essentially matches the notion of atomism that classical physics is assumed to implicitly uphold. Howard (1989, p. 226), drawing inspiration from Einstein's (1948/1971) views on physical reality, offers a formulation of the principle of separability, as a fundamental metaphysical constraint on physical systems and their associated states, along the following lines:

> *Separability Principle*: Each physical system possesses its own distinct, separate state — which determines its qualitative, intrinsic properties — and the whole state of any compound system is completely determined by the separate states of its subsystems and their spatiotemporal relations.

Two points of consideration are here in order: Firstly, the aforementioned characterization of the separability principle implies in a natural manner a supervenience claim; for, if the whole state of a compound system is completely determined by the separate states of its subsystems, then the whole state necessarily supervenes on the separate states. Secondly, it is compatible with Lewis' conception of Humean supervenience. As one may recall, Lewis' version of Humean supervenience consists in conceiving the physical properties on which everything else supervenes as properties of points of space-time. In this case, separability about states of physical systems follows straightforwardly. For, if one starts from Humean supervenience, the contents of any two spatiotemporally separated points can be considered to constitute separate physical systems, the joint state of which is fully determined by the local properties that are attributed to each of these points.

Both doctrines of separability and Humean supervenience are inspired by classical physics, be it point-like analytic or field theoretic. With respect to the latter, for instance, the essential characteristic of any classical field theory, regardless of its physical content and assumed mathematical structure, is that the values of fundamental parameters of a field are well defined at every point of the underlying manifold. In the case, for example, of general relativity (*qua* classical field theory), exhaustive specification of the ten independent components of the metric tensor at each point within a given region of the space-time manifold, completely determines the gravitational field in that region. In this sense, the total existence of a field in a given region is contained in its parts, namely, its space-time points. Thus, in attributing physical reality to point-values of basic field parameters, a field theory proceeds by tacitly assuming that a physical state is ascribed to each point of the manifold, and this state determines the local properties of this point-system. Furthermore, the compound state of any set of such point-systems is completely determined by the individual states of its constituents. Hence, classical field theories necessarily satisfy the separability principle and are therefore subjected to the doctrine of Humean supervenience.[5]

Similar considerations arise through the particle-theoretic viewpoint of classical physics. Within the framework of point-like analytic mechanics, the state of a compound system S consisting of N point particles is specified by considering all pairs $\{q_{3N}(t), p_{3N}(t)\}$ of the physical quantities of position and momentum of the individual particles instantiated at distinct space-time points. Hence, at any temporal moment t, the individual pure state of S consists of the N-tuple $\omega = (\omega_1, \omega_2, ... , \omega_N)$, where $\{\omega_i\} = \{q_i, p_i\}$ are the pure states of its constituent subsystems. Consequently, every property the compound system S has at time t, if encoded in $\omega$, is determined by $\{\omega_i\}$. For example, any classical physical quantities (such as mass, momentum, angular momentum, kinetic energy, center of mass motion, gravitational potential energy, etc.) pertaining to the overall system are determined in terms of the corresponding local quantities of its parts. They either constitute direct sums or ordinary functional relations (whose values are well-specified at each space-time point) of the relevant quantities of the subsystems. Thus, they are wholly determined by, and hence supervenient on, the subsystem states. In this respect, every classical compound system S is not only admissible to the dictates of the doctrine of Humean supervenience, but, most importantly, it is separable. As such, it is necessarily Humean supervenient upon the joint existence of its relata, namely, its constituent subsystems $S_1, S_2, ..., S_N$. Consequently, every qualitative, intrinsic property and relation pertaining to a classical system S, considered as a whole, is supervenient upon the qualitative, intrinsic properties of its basic related parts, in conjunction with their spatio-temporal relations.

What exactly means, however, for a relation to supervene upon the qualitative, intrinsic (and therefore non-relational) properties of the relata to which it refers? Cleland (1984, p. 25) provides a formal characterization of a supervenient relation in modal terms that captures the essential gist of our preceding analysis of supervenience between properties as a 'dependence – determination' relationship:

A dyadic relation *R* is *supervenient* upon a determinable non-relational attribute *P* if and only if:

(1) $(\forall x, y) \sim \Diamond \{R(x, y)$ and there are no determinate attributes $P_i$ and $P_j$ of determinable kind $P$ such that $P_i(x)$ and $P_j(y)\}$;

(2) $(\forall x, y) \{R(x, y) \supset$ there are determinate attributes $P_i$ and $P_j$ of determinable kind $P$ such that $P_i(x)$ and $P_j(y)$ and $(\forall x, y) [(P_i(x)$ and $P_j(y)) \supset R(x, y)]\}$.[6]

The above two conditions can be explicated as follows: If relation *R* is genuinely supervenient upon *P*, then condition (1) implies that *R* cannot possibly appear in the absence of each of its relata instancing the requisite property *P*, whereas condition (2) adds that there must exist one or more pairs of determinate non-relational properties (of kind *P*) whose exemplification alone is sufficient to guarantee the appearance of *R*.

Stating these two necessary and sufficient conditions as a characterization of supervenience, enables Cleland to distinguish between two kinds of 'non-supervenience' in terms of strongly and weakly non-supervenient relations:

*Strong non-supervenience*: a relation is strongly non-supervenient if the appearance of this relation is neither dependent upon nor determined by non-relational properties of its relata. In other words, a relation is strongly non-supervenient if and only if it violates both conditions (1) and (2). That is to say, if two entities bear a strongly non-supervenient relation to each other, then necessarily should possess no intrinsic properties of a certain kind instancing the relation in question.

*Weak non-supervenience*: a relation is weakly non-supervenient if the appearance of this relation is dependent upon the instantiation of non-relational properties of each of its relata, but the latter are not sufficient to determine the relation in question. Accordingly, a relation is weakly non-supervenient if and only if it satisfies condition (1), but violates condition (2).

It should be observed that Cleland's formulation of a supervenient relation and its negation is independent of any predilections concerning the conceptual foundations of contemporary physics. On the other hand, French (1989), inspired by Teller (1986), introduced Cleland's conditions into the interpretation of quantum theory as a reasonable alternative to understanding the puzzling EPR-entangled correlations in terms of non-supervenient relations. In the next section, Cleland's conditions are projected into the context of quantum nonseparability as a means of rigorously evaluating the compatibility status of Humean supervenience against the background of foundational physics.

## 4. Quantum Entangled Relations and Humean Supervenience

Classical physics, as already demonstrated, provides strong motivation to Lewis' thesis of Humean supervenience, at least, to the extent that the latter incorporates at its foundation the aforementioned separability principle of Section 3. The notion of separability has been regarded within the framework of classical physics as a principal condition of our conception of the world, a condition that characterizes all our thinking in acknowledging the physical identity of distant things, the "mutually independent existence (the 'being thus')" of spatiotemporally separated systems (Einstein 1948/1971, p.169). The principle of separability delimits in essence the fact — upon which the whole classical physics is implicitly founded — that any compound physical system of a classical universe can be conceived of as consisting of *separable*, *individual* parts interacting by means of forces, which are encoded in the Hamiltonian function of the overall system, and that the state of the latter (in the limiting case, the physical state of the world itself) is completely determined by the intrinsic physical properties pertaining to each one of these parts and their spatiotemporal relations. In contradistinction, standard quantum mechanics systematically violates the conception of separability that classical physics accustomed us to consider as valid.[7] From a formal point of view, the source of its defiance is due to the tensor-product structure of a compound Hilbert space and the quantum mechanical principle of the superposition of states, incorporating a kind of objective indefiniteness for the numerical values of any physical quantity belonging to a superposed state.

### *4.1. Strongly Non-Supervenient Relations*
As a means of explicating the preceding factors in relation to Lewis' doctrine of Humean supervenience, let us consider the simplest possible case of a compound system S consisting of a pair of subsystems $S_1$ and $S_2$ with corresponding Hilbert spaces $H_1$ and $H_2$. Naturally, subsystems $S_1$ and $S_2$, in forming system S, have interacted by means of forces at some time $t_0$ and suppose that at times $t > t_0$ they are spatially separated. Then, any pure state W of the compound system S can be expressed in the tensor-product Hilbert space $H = H_1 \otimes H_2$ in the Schmidt form

$$W = P_{|\Psi\rangle} = |\Psi\rangle\langle\Psi| = \sum_i c_i (|\psi_i\rangle \otimes |\varphi_i\rangle), \qquad \| |\Psi\rangle \|^2 = \sum_i |c_i|^2 = 1, \quad (1)$$

where $\{|\psi_i\rangle\}$ and $\{|\varphi_i\rangle\}$ are orthonormal vector bases in $H_1$ (of $S_1$) and $H_2$ (of $S_2$), respectively.

If there is just one term in the W-representation of Eq. (1), i.e., if $|c_i| = 1$, the state $W = |\psi\rangle \otimes |\varphi\rangle$ of the compound system forms a product state: a state that can always be decomposed into a single tensor-product of an $S_1$-state and an $S_2$-state. In this circumstance, each subsystem of the compound system possesses a separable and well-defined state, so that the state of the overall system consists of nothing but the logical sum of the subsystem states in consonance with the separability principle of Section 1. This is the only highly particular as well as idealized case in which a separability principle holds in quantum mechanics. For, even if a compound system at a given temporal instant t is appropriately described by a product state — $W(t) = |\psi_{(t)}\rangle \otimes |\varphi_{(t)}\rangle$ — the preservation of its identity under the system's natural time evolution — $W(t_2) = U(t_2-t_1) W(t_1)$, for all $t \in \mathbb{R}$ — implies that the Hamiltonian *H* (i.e., the energy operator of the system) should be decomposed into the direct sum of the subsystem Hamiltonians — $H = H_1 \otimes I_2 + I_1 \otimes H_2$ — and this is precisely the condition of *no interaction* between $S_1$ and $S_2$ (e.g., Blank et al. 1994, Ch. 11). Obviously, in such a case, subsystems $S_1$ and $S_2$ behave in an entirely uncorrelated and independent manner. Correlations, even of a probabilistic nature, among any physical quantities corresponding to the two subsystems are simply non existent, since for any two observables $A_1$ and $A_2$ pertaining to $S_1$ and $S_2$, respectively, the probability distributions of $A_1$ and of $A_2$ are disconnected: $\text{Tr}(A_1 \otimes A_2)(|\psi\rangle \otimes |\varphi\rangle) = \text{Tr}(A_1 |\psi\rangle) \cdot \text{Tr}(A_2 |\varphi\rangle)$.

If, however, there appear more than one term in the W-representation of the compound system, i.e., if $|c_i| < 1$, then there exist entangled correlations (of the well-known EPR-type) between subsystems $S_1$ and $S_2$. It can be shown in this case, as already indicated by Schrödinger (1935/1983, p. 161), that there are no subsystem states $|\xi\rangle$ ($\forall |\xi\rangle \in H_1$) and $|\chi\rangle$ ($\forall |\chi\rangle \in H_2$) such that W is equivalent to the conjoined attribution of $|\xi\rangle$ to subsystem $S_1$ and $|\chi\rangle$ to subsystem $S_2$, i.e., $W \neq |\xi\rangle \otimes |\chi\rangle$. Thus, when a compound system, such as S, is in an entangled state W, namely a superposition of pure states of tensor-product forms, neither subsystem $S_1$ by itself nor subsystem $S_2$ by itself is associated with an individual pure state. The normalized unit vectors $|\psi_i\rangle$, $|\varphi_i\rangle$ belonging to the Hilbert space of either subsystem are not eigenstates of the overall state W. If descriptions of physical systems are restricted to the state vector assignment of states, then, strictly speaking, subsystems $S_1$ and $S_2$ have no states at all, even when $S_1$ and $S_2$ are spatially separated. Only the compound system is assigned a definite (nonseparable) pure state W, represented appropriately by a state vector $|\Psi\rangle$ in the tensor-product Hilbert space of S. Maximal determination of the whole system, therefore, does not allow the possibility of acquiring maximal specification of its component parts, a circumstance with no precedence in classical physics.

Since on the state vector ascription of states, neither subsystem $S_1$ nor subsystem $S_2$ has a state vector in S, it is apparent that the state W of the compound system cannot be reproduced on the basis that neither part has a well-defined individual state, namely, a pure state. Thus, the separability principle of Section 3 is violated and likewise Lewis' version of Humean supervenience does fail. The entangled state W represents global properties for the whole system S that are neither *dependent* upon nor *determined* by any properties of its parts. In fact, in any case of quantum entanglement, conceived as a relation among the constitutive parts of a quantum whole, there exist no qualitative, non-relational properties of the parts whose exemplification is sufficient to guarantee the appearance of entanglement. In such a case both supervenience conditions (1) and (2) are clearly violated; for not only the entangled relation pertaining to compound system S is not guaranteed by the exemplification of qualitative, non-relational properties of subsystems $S_1$ and $S_2$, thus violating supervenience condition (2), but neither of the subsystems $S_1$ and $S_2$, being the related parts of the entangled relation, exemplify any qualitative, non-relational properties at all, thus violating supervenience condition (1). Hence, it may be asserted that the relation of quantum entanglement is *inherent* to the compound system S, as a whole, exhibiting strongly non-supervenient behavior with respect to the relata $S_1$ and $S_2$. In this respect, Lewis' thesis of Humean supervenience fails on, at least, two counts: firstly, there exist non-supervenient relations beyond the spatiotemporal ones, namely quantum mechanical entangled relations, and, secondly, in considering any such relation, there exist no non-relational states (or properties) of its related parts.

As a means of exemplifying the preceding points, let us consider an important class of compound systems that form the prototype of quantum entangled systems, namely, spin-singlet pairs. Let then S be a compound system consisting of a pair $(S_1, S_2)$ of spin-1/2 particles in the singlet state

$$W_S = 1/\sqrt{2} \; \{|\psi_+>_1 \otimes |\varphi_->_2 \; - \; |\psi_->_1 \otimes |\varphi_+>_2\}, \quad (2)$$

where $\{|\psi_\pm>_1\}$ and $\{|\varphi_\pm>_2\}$ are spin-orthonormal bases in the two-dimensional Hilbert spaces $H_1$ and $H_2$ associated with $S_1$ and $S_2$, respectively. As well known, in such a case, it is quantum mechanically predicted and experimentally confirmed that the spin components of $S_1$ and of $S_2$ have always opposite spin orientations; they are perfectly anticorrelated. Whenever the spin component of $S_1$ along a given direction is found to be +1/2 ℏ (correspondingly −1/2 ℏ), then the spin component of $S_2$ along the same direction must necessarily be found to be −1/2 ℏ (correspondingly +1/2 ℏ), and conversely. From a physical point of view, this derives from the interference (the definite phase interrelations) with which the subsystem unit vectors $|\psi_\pm>$ and $|\varphi_\pm>$ — or, more precisely, the two product states $|\psi_+>_1\otimes|\varphi_->_2$, $|\psi_->_1\otimes|\varphi_+>_2$ — are combined within $W_S$. This, in turn, leads not only to the subsystem interdependence of the type described above, but also to conservation of the total angular momentum for the pair $(S_1, S_2)$ of spin-1/2 particles, and thus to the property of definite total spin of value zero for the compound system S.

The latter is a *holistic* property of S: it is not determined by any physical properties of its subsystems $S_1$, $S_2$ considered individually. Specifically, the property of S 'having total spin zero' is strongly non-supervenient on the spin properties of $S_1$ and of $S_2$, since neither $S_1$ nor $S_2$ has any definite spin in the singlet state $W_S$. Moreover, the probability distributions concerning spin components of $S_1$ and of $S_2$ along some one direction do not ensure, with probability one, S's possession of this property. Neither the latter could be understood or accounted for by the possibility (that an adherent of Humean supervenience may favor) of treating $S_1$ and $S_2$ separately at the expense of postulating a relation between them as to the effect of their spin components 'being perfectly anticorrelated'. For, while 'having total spin zero' is an intrinsic physical property of the compound system S in the nonseparable state $W_S$, the assumed relation is not an intrinsic physical relation that $S_1$ and $S_2$ may have in and of themselves. That is, although the relation of perfect anticorrelation is encoded within state $W_S$, ascribing this relation to individual parts of a system is not tantamount to being in state $W_S$. The relation of perfect anticorrelation is inherent to the entangled state $W_S$ itself which refers directly to the whole system. The entangled correlations between $S_1$ and $S_2$ just do not supervene upon any (intrinsic or extrinsic) properties of the subsystem parts taken separately.

It may seem odd to consider non-supervenient relations holding between non-individuatable relata. However, the important point to be noticed is that within an entangled quantum system there is no individual pure state for a component subsystem alone. Within $W_S$ neither subsystem $S_1$ nor subsystem $S_2$ acquire individual independent existence. In considering any entangled compound system, the nature and properties of component parts may only be determined from their 'role' — the forming pattern of the inseparable web of relations — within the whole. Here, the part-whole relationship appears as complementary: the part is made 'manifest' through the whole, while the whole can only be 'inferred' via the interdependent behavior of its parts (e.g., Karakostas 2007). Thus, in the example under consideration, the property of total spin of the whole in the singlet state $W_S$ does indicate the way in which the parts are related with respect to spin, although neither part possesses a definite numerical value of spin in any direction in distinction from the other one. And it is *only* the property of the total spin of the whole that contains *all* that can be said about the spin properties of the parts, because it is only the entangled state of the whole that contains the correlations among the spin probability distributions pertaining to the parts.[8] Consequently, the part-whole reduction with respect to the property of total spin zero in $W_S$ has failed: the latter property, whereas characterizes the whole system, is not supervenient upon — let alone being reducible to — any properties of its constituent parts. Exactly the same holds for the properties of total momentum and relative distance of the overall system S with respect to corresponding local properties of the parts. Analogous considerations, of course, to the aforementioned paradigmatic case of the spin-singlet pair of particles apply to any case of quantum entanglement. Entanglement need not be of maximal anticorrelation, as in the example of the singlet state.[9] It does neither have to be confined to states of quantum systems of the same kind; entanglement reaches in principle the states of all compound quantum systems.

The generic phenomenon of quantum entanglement and the associated conception of quantum nonseparability cast severe doubts on the applicability of the doctrine of Humean supervenience. In fact, the non-supervenient relations of entanglement among the parts of a quantum whole imply a reversal of Lewis' thesis of Humean supervenience: for, there exist properties pertaining to any entangled quantum system which, in a clearly specifiable sense, characterize the whole system but are neither supervenient upon nor reducible to or derived from any combination of local properties of its parts. On the contrary, it is only the entangled state of the whole system which completely determines the local properties of its subsystem parts and their relations (to the extent that these are determined at all; see further Section 4.2). If Lewis' version of Humean supervenience held within quantum mechanics, one could analyze the entangled state of a compound system into local physical states of the component parts taken separately in conjunction with the spatiotemporal relations among the parts. In such a case, however, the state of the compound system would be a product state, in flagrant contradiction with the initial assumption of entanglement. Evidently, Lewis' thesis of Humean supervenience — proclaiming that the spatiotemporal arrangement of local intrinsic qualities provides a subvenient basis upon which all else supervenes — comes into conflict with contemporary physics. Within the framework of quantum mechanics, even the core presupposition of Lewis' Humean assumption that there exist intrinsically local entities furnishing the bearers of the fundamental properties and relations comes into question. For the nonseparable character of the behavior of an entangled quantum system precludes in a novel way the possibility of describing its component subsystems as well-defined individuals, each with its own pure state or pre-determined physical properties. Upon any case of quantum entanglement, it is not permissible to consider the parts of a quantum whole as self-autonomous, intrinsically defined individual entities. In fact, as considered immediately below, whenever the pure entangled state of a compound system is decomposed in order to represent well-defined subsystems, the effect can only extent up to a representation in terms of statistical (reduced) states of those subsystems.

### *4.2. Weakly Non-Supervenient Relations*

In view, therefore, of the radical violation of Humean supervenience on the state vector ascription of quantum states, it is interesting to inquire whether the assignment of statistical states to component subsystems, represented by non idempotent density operators, restores a notion of Humean supervenience into quantum theory? The question, however, even according to such a circumstance of physical possibility, is answered strictly in the negative.

The clearest way to establish this, for present purposes, is by regarding again as the state W of a compound system the singlet state of a pair of spin-1/2 particles $(S_1, S_2)$ in the familiar development

$$W_S = 1/\sqrt{2}\ \{|\psi_+\rangle_1 \otimes |\varphi_-\rangle_2 - |\psi_-\rangle_1 \otimes |\varphi_+\rangle_2\}. \quad (3)$$

Observe, in consonance with the considerations of Section 2.1, that neither particle $S_1$ nor particle $S_2$ can be represented in $W_S$ of Eq. (3) by a state vector. However, each particle may be assigned a state, albeit a reduced state, that is given by the partial trace of the density operator $W_S$ of the compound system. Recall that the reduced state of each particle arises by 'tracing over' the degrees of freedom associated with the Hilbert space representation of the partner particle. Hence, the following density operators

$$\hat{W}_1 = 1/2\ P_{|\psi_+\rangle} + 1/2\ P_{|\psi_-\rangle} \quad \text{and} \quad \hat{W}_2 = 1/2\ P_{|\varphi_+\rangle} + 1/2\ P_{|\varphi_-\rangle} \quad (4)$$

represent the reduced ('unpolarized') states of spin-1/2 particles $S_1$ and $S_2$, respectively, in state $W_S$.[10]

It is not hard to show, however, that the component states $\hat{W}_1$ and $\hat{W}_2$ of (4) could be identical if derived from a compound state W′ that would correspond to the triplet state

$$W' = 1/\sqrt{2}\ \{|\psi_+\rangle \otimes |\varphi_-\rangle + |\psi_-\rangle \otimes |\varphi_+\rangle\}, \quad (5)$$

or to the following pure states (proviso the sign)

$$W'' = 1/\sqrt{2} \ \{|\psi_+\rangle \otimes |\varphi_+\rangle \ \pm \ |\psi_-\rangle \otimes |\varphi_-\rangle\} \quad (6)$$

that yield, in general, different predictions than W or W' does for certain spin measurements of both $S_1$ and $S_2$ along a given direction; or they still could be identical if derived from the mixed state

$$W''' = 1/2 \ \{|\psi_+\rangle \otimes |\varphi_-\rangle \ + \ |\psi_-\rangle \otimes |\varphi_+\rangle\}. \quad (7)$$

Thus, given the states $\hat{W}_1$ and $\hat{W}_2$ of subsystems $S_1$ and $S_2$, respectively, the compound state could equally well be either W or W', W'', W''' or in numerous other variations of them. There exists a many-to-one mapping of the subsystem non-pure states to the state of the whole system.[11] Accordingly, the specification of the compound system remains indefinite. For, the compound state $W_S$ contains correlations between subsystems $S_1$ and $S_2$ that the reduced states $\hat{W}_1$ and $\hat{W}_2$ do not contain. The sort of correlations that is missing corresponds, from a formal point of view, to the tracing out in the specification, for instance, of $\hat{W}_1$ of what might be known about the state of subsystem $S_2$ and about its connection with subsystem $S_1$. It is evident, therefore, that at the level of description in terms of a reduced state, the information obtained by considering, even simultaneously, the two subsystems does not permit the reconstruction of the pure state of the whole system, i.e., $W \neq \hat{W}_1 \otimes \hat{W}_2$. In this sense, it may be said that the 'whole' (e.g., system S) is, in a non-trivial way, more than the combination of its 'parts' (e.g., subsystems $S_1$ and $S_2$), this being the case even when these parts occupy remote regions of space however far apart. Hence, on the density operator ascription of states, whereas each subsystem possesses a local state independently of the other, still the state of the whole system fails to be exhausted by a specification of the states of the parts and their spatiotemporal relations. Consequently, the separability principle of Section 3 is violated.

If the preceding state of affairs be translated in terms of state-dependent properties, it immediately follows that the exemplification of non-relational, qualitative properties pertaining to $\hat{W}_1$ and $\hat{W}_2$ is not sufficient to guarantee the exemplification of the relational properties incorporated in $W_S$. In such a circumstance, supervenience condition (1) is satisfied, whereas supervenience condition (2) is clearly violated. Thus, even on the statistical assignment of states to component subsystems of a quantum whole, one is still confronted with weakly non-supervenient relations.

The difficulty facing Lewis' doctrine of Humean supervenience is now unavoidable. Consider two pairs of spin-1/2 particles, one in the singlet and the other in the triplet state, represented by Eqs. (3) and (5), respectively, such that the spatio-temporal relations within each pair are identical. Each particle of either pair in the singlet or triplet state can now be assigned its own local state, a reduced statistical state in accordance with the expressions of Eq. (4). It is important to note that both singlet and triplet states assign the same spin probability distributions for each component particle. No local spin-measurements performed on a component particle of a pair in the singlet state can distinguish it from a component of a pair in the triplet. For, the statistical states $\hat{W}_1$ and $\hat{W}_2$ assigned to component particles in the singlet state are *identical* to the statistical states assigned to component particles in the triplet state; they are equiprobable 'mixtures' of the states $|\psi_\pm\rangle$ and $|\varphi_\pm\rangle$. Consequently, if Lewis' version of Humean supervenience holds, then, since each component particle in the singlet state is in precisely the same spin state as each component in the triplet state, and since the spatio-temporal relations between the component particles of each pair are identical, the singlet state would have to be identical to the triplet state in Lewis' terms. These are, however, significantly different. The singlet state gives rise to a global property for total spin of eigenvalue zero, whereas, the triplet state predicts with certainty the eigenvalue result $2\hbar^2$. Both values may be verified by joint measurements on the two-component system for the total (squared) spin operator in the corresponding states. Thus, there is a difference in global properties to which no difference in the local properties of the component parts corresponds. Hence, Lewis' thesis of Humean supervenience fails.

### *4.3. The Status of the Recombination Principle in Light of Quantum Entanglement*

Likewise fails his recombination principle. The latter constitutes a re-expression of the Humean denial of necessary connections between distinct existences. In short, Lewis' recombination principle requires that ''anything can coexist with anything, and which thereby prohibits a necessary connection between the intrinsic character of a thing and the intrinsic character of distinct things with which it coexists'' (Lewis 1986b, p. 181, see also pp. 86-92). To realize the scope of Lewis' recombination principle, it is useful to recall that his metaphysical doctrine of Humean supervenience conceives the world as being consisted of an enormous number of local matters of particular fact. The world is composed of *discrete* events, like a vast mosaic, ''just one little thing and then another'' with no tie or connection between one and the other. Each event is wholly *self-contained*, all its properties are entirely *intrinsic*; its nature is *exhausted* within the spatio-temporal boundary that occupies. Of course, each event may coexist alongside with any other particular or particulars to form any pattern. In Lewis' vision of the world, however, even if a sequence or a combination of events presents a recognizable pattern, the individual event has a nature and existence *independent* of the pattern they formed. Each event of this pattern is causally inert,[12] thus incapable of reaching out beyond itself to any other event. It is a particular whose nature is completely independent of anything around it. Hence, there are *no* necessary connections between an individual event and any of its neighbours or other events in the pattern. The latter simply consists of the mereological sum of all particulars that are part of it. The same events — by virtue of their mutual independent existence — could be *recombined* into any order and arrangement so as to form a completely different pattern. At any circumstance, the events, and their spatio-temporal relations among each other, constitute the whole of the objectively real world. From the perspective of Lewis' metaphysic, if one is able to determine the intrinsic qualities of particular events or atomic objects in space and time, then one can describe the world completely.

Quantum mechanics, however, is not in conformity with Lewis' atomistic metaphysical picture that depicts a world of self-contained, unconnected particulars that exist independently of each other. As we have extensively argued, within the quantum theoretical context, the consideration of physical reality cannot be comprehended as the sum of its parts in conjunction with the spatiotemporal relations among the parts, since the quantum whole provides the framework for the existence of the parts. In contradistinction to Lewis' conception, whole and parts, now, mutually influence and specify each other; the parts are adjusted by the whole, whereas, in turn, the whole depends on the interconnectedness of the parts. Upon any case of quantum entanglement, the interrelation between the parts cannot be disclosed in an analysis of the parts that takes no account of the entangled connection of the whole. As already shown, their entangled relation does not supervene upon any intrinsic or relational properties of the parts taken separately. This is indeed the feature which makes the quantum theory go beyond any mechanistic or atomistic thinking. In Lewis' view, given any compound physical system, the dynamics of the whole is reducible to the properties of the parts. In quantum mechanics the situation is actually reversed; due to the genuinely nonseparable structure of quantum theory, the identity and properties of the parts can ultimately be accounted only in terms of the dynamics of the whole.[13] In a truly nonseparable physical system, as in an entangled quantum system, the part does acquire a different identification within the whole from what it does outside the whole, in its own 'isolated', separate state (see esp. Section 4.1). Thus, for instance, no isolated spin-1/2 particle, e.g. a 'free' or 'bare' electron, can be identified with the spin state of either member of a pair of electrons in the singlet state, since in this situation any spin state can be specified only at the level of the overall system. When in the singlet state, there is simply no individual spin state for a component particle alone, unless explicit reference is made to the partner particle via the total information contained in the compound state. Consequently, any spin state of either particle is fixed only through the interconnected web of entangled relations among the particles. Hence, the spin property of either particle, when in an entangled state, cannot stand alone, unconnected to the whole pattern and unaffected by anything else, in flagrant contradiction with Lewis' suggestion concerning his recombination principle.

### *4.4. Debugging Humean Supervenience ?*

Lewis (1986a, pp. x-xi) acknowledges that quantum theory may pose a threat to Humean supervenience:

> … it just might be that Humean supervenience is true, but our best physics is dead wrong in its inventory of the qualities. Maybe, but I doubt it. … what I uphold is not so much the truth of Humean supervenience as the *tenability* of it. If physics itself were to teach me that it is false, I wouldn't grieve. That might happen: maybe the lesson of Bell's theorem is exactly that there are *physical* entities which are unlocalized, and which might therefore make a difference between worlds … that match perfectly in their arrangements of local qualities. Maybe so. I'm ready to believe it. But I am not ready to take lessons in ontology from quantum physics as it now is. First I must see how it looks when it is purified of instrumentalist frivolity, and dares to say something not just about pointer readings but about the constitution of the world; and when it is purified of doublethinking deviant logic; and — most all — when it is purified of supernatural tales about the power of the observant mind to make things jump. If, after all that, it still teaches nonlocality, I shall submit willingly to the best of authority.

Lewis (1994, p. 474) in addition offers the following qualification regarding his defence of the thesis of Humean supervenience:

> The point of defending Humean Supervenience is not to support reactionary physics, but rather to resist philosophical arguments that there are more things in heaven and earth than physics has dreamt of. Therefore if I defend the *philosophical* tenability of Humean Supervenience, that defence can doubtless be adapted to whatever better supervenience thesis may emerge from better physics.

It should be observed in relation to Lewis' first quote that quantum theory can no longer be viewed as an awkward and annoying phase through which physical science is progressing, only later surely to settle down as 'philosophically more respectable'. Quantum theory has reached a degree of maturity that simply cannot be ignored on the basis of a priori philosophical expectations. In the course of the multifaceted investigations of the conceptual foundations of quantum mechanics, there has been developed a variety of formulations of the theory in which neither the concept of 'measurement' nor the concept of 'the observant mind' bear any fundamental role. Approaches of this kind involve, for instance, the consistent histories formulation of Griffiths (1984) and Omnès (1992), the decoherence theory of Zurek (1982, 1991), the stochastic dynamical localization model of Ghirardi, Rimini, Weber (1986) and Pearle (1989), or even Everett's (1957) relative-state formulation. All these theories posit fundamental nonseparable physical states of affairs. Nonseparability constitutes a structural feature of quantum mechanics that distinctly marks its entire departure from classical lines of thought. If, therefore, present day quantum theory, or any extension of it, is part of a true description of the world, then Lewis' doctrine of Humean supervenience cannot be regarded as a reliable code of the nature of the physical world and its contents. Quantum theory implies in fact, as extensively argued, an irrevocable failure of Humean supervenience. For, it shows that in considering any compound quantum system, there simply exist no intrinsic local properties of the entangled systems involved which could form a basis on which the relation of entanglement could supervene. Acknowledging the generic character of the latter at the microphysical level, no proper foundation can be established within contemporary physics for formulating the doctrine of Humean supervenience. Consequently, Lewis' qualification that his defence of Humean supervenience "can doubtless be adapted to whatever better supervenience thesis may emerge from better physics", is clearly not the case. The thesis of Humean supervenience is ill-formed; it relies upon a false presupposition, namely separability.

Neither the violation of the separability principle, as established by quantum theory, revives the existence of "more things in heaven and earth than physics has dreamt of". On the contrary, quantum nonseparability only implies that there are more fundamental relations dictated by physics — namely, quantum mechanical entangled relations — than Humean supervenience can possibly accommodate. The physical existence of quantum entangled relations, as well as their empirical confirmation,[14] undeniably shows that what there is in the world is more tightly intertwined than just by spatiotemporal relations among separately existing entities that are localized at space-time points. In contrast to the thesis of Humean supervenience, therefore, relations of quantum entanglement are proven to be at least as fundamental as spatiotemporal relations in unifying a world.

## 5. CONCLUDING REMARKS

The previously presented arguments establish the existence of non-spatiotemporal, non-supervenient physical relations, undermining, thereby, the metaphysical doctrine of Humean supervenience. For, contrary to Lewis' contentions, contemporary microphysics posits irreducible holistic relations that do not supervene upon a spatiotemporal arrangement of Humean properties. Specifically, *any* relation of quantum entanglement among the parts of a compound system endows the overall system with properties which are neither *reducible to* nor *dependent on* or *supervenient upon* any (intrinsic or extrinsic) properties that can possibly be attributed to each of its parts. Hence, due to the all-pervasive phenomenon of quantum entanglement at the microphysical level, the functioning of the physical world cannot just be reduced to that of its constituent parts in conjunction with the spatiotemporal relations among the parts. In this respect, the assumption of ontological reductionism, as expressed in the thesis of Humean supervenience, can no longer be accepted as a true precept of the nature of the physical world and its contents.

The foregoing considerations also illustrate a more general lesson for metaphysics. Although it may seem quite plausible that no physical relation can *fully* determine the qualitative, intrinsic properties (if any) of its relata, the metaphysical doctrine of Humean supervenience implies that a prior spatiotemporal determination of them can fix *any* relation holding among the relata. This alleged possibility has ultimately been shown to be founded on intuitive grounds or a priori philosophical generalization. For, the undeniable existence of generic non-supervenient relations within fundamental physics induces one to admit that certain relations (on a par with certain qualitative intrinsic properties) are basic constituents of the world. Weak or strong non-supervenience of specific relations upon non-relational facts clearly indicates that such relations have a physical reality on their own. They should be admitted, therefore, into our natural ontology as genuine irreducible elements. Consequently, a metaphysics of relations of a moderate kind — far from being deemed as paradoxical, as frequently is the case in current philosophical thought — ought to be acknowledged as an indispensable part of our understanding of the natural world at a fundamental level.

## NOTES

[1] Teller (1989) himself employs the term 'particularism' to characterize the notion of atomism that is implicit in classical physics. This usage of the term, however, is not entirely appropriate, since the concept of 'particularism' is normally viewed as a means of denying the existence of 'universals'. See, for instance, Simons (1994, p. 557).

[2] There is no a unanimously accepted definition either of the notion of an intrinsic property, or the conception of a qualitative property. For an introductory survey on this matter, see Weatherson (2002). Various suggestions for defining intrinsic properties are found, for instance, in Lewis (1986, pp. 262-266) and, more recently, in Vallentyne (1997). For helpful distinctions between 'intrinsic' and 'extrinsic' properties, see Humberstone (1996). Compare with Langton and Lewis (1998).

[3] Both atomism and holism are less a doctrine, than a class of doctrines expanding from ontological, explanatory or methodological qualifications to biological, social, mental, linguistic or semantic considerations. The connotation 'physical' is used precisely to restrict analysis to physical science with respect to atomism/holism issue.

[4] Characteristic is the following claim of Hume (1975/1748, p. 74) in his "Enquiry Concerning Human Understanding", arguing against the epistemic accessibility or, according to the customary interpretation of Hume, against the reality of necessary connections in nature:

> …upon the whole, there appears not, throughout all nature, any one instance of connexion which is conceivable by us. All events seem entirely loose and separate. One event follows another; but we never can observe any tie between them. They seem *conjoined*, but never *connected*. And as we can have no idea of any thing which never appeared to our outward sense or inward sentiment, the necessary conclusion *seems* to be that we have no idea of connexion or power at all, and that these words are absolutely without any meaning, when employed either in philosophical reasonings or common life.

[5] It is of no coincidence, in this respect, that Lewis' examples of 'Humean' properties, namely, intrinsic properties that require no more than a spatiotemporal point to be instatntiated, are the values of the electromagnetic and gravitational fields (see also Loewer 1996, p. 102).

[6] Two points of clarification should be noted: First, the symbols " " and "◊" denote the modal operators for necessity and possibility. Thus, " P" should be read "it is necessary that P" and "◊ P" should be read "it is possible that P". Second, the differentiation between 'determinable' and 'determinate' should be understood in the sense of a determinable attribute of a certain kind — such as having spin — and a determinate attribute — such as having spin of, say, 1/2 along a certain direction.

[7] In this work we shall not consider in any detail alternative interpretations to Hilbert-space quantum mechanics as, for instance, Bohm's ontological or causal interpretation.

[8] In this connection see Esfeld (2004). Also Rovelli (1996) and Mermin (1998) highlight the significance of correlations as compared to that of correlata.

[9] It is well known that spin-singlet correlations violate Bell's inequalities. We note in this connection the interesting result of Gisin (1991), Popescu and Rohrlich (1992) that for *any* entangled state of a two-component system there is a proper choice of pairs of observables whose correlations do violate Bell's inequality.

[10] It is worthy to note that the non-purity of the subsystem states of Eq. (4) arises as a restriction of the overall pure state $W_S$ of the entangled system to the observables pertaining to a component subsystem. Any subsystem state in this situation is exclusively defined at the level of the whole system; there is no individual state for a component subsystem alone. For, there exists no justification in regarding the prescribed reduced states of Eq. (4) as being associated with any specific ensemble of pure states with corresponding eigenvalue probabilities. In this respect, the reference of a reduced state is only of a statistical, epistemic nature. It simply reflects the statistics that may be derived by a series of local measurements performed on a given component subsystem.

[11] Hughston et al. (1993) provide a constructive classification of all discrete ensembles of pure quantum states that correspond to a fixed density operator.

[12] In Lewis' philosophical scheme, the notion of causation should be understood in a broadly neo Humean way as a *contingent* relation between 'distinct existences'. Causal relationships are, roughly speaking, nothing but patterns which supervene on a point-by-point distribution of properties.

[13] It should be noted that the so-called invariant or state-independent properties — like 'rest-mass', 'charge' and 'spin' — of elementary objects-systems can only characterize a certain class of objects; they can only specify a certain sort of particles, e.g., electrons, protons, neutrons, etc. They are not sufficient, however, for determining a member of the class as an individual object, distinct from other members within the same class, that is, from other objects having the same state-independent properties. Thus, an 'electron', for instance,

could not be of the particle-kind of 'electrons' without fixed, state-independent properties of 'mass' and 'spin', but these in no way suffice for distinguishing it from other similar particles or for 'individuating' it in any particular physical situation. For a detailed treatment of this point, see, for example, Castellani (1999).

[14]See, for instance, Aspect et al. (1982); also the relatively recent result of Tittel et al. (1998).